\documentclass[aps,prl,reprint,
superscriptaddress,
mathptmx,
inputenc,
fontenc,
graphicx,
siunitx,
]{revtex4-2}
\usepackage[USenglish]{babel} 
\usepackage{mathptmx}
\usepackage[utf8]{inputenc} 
\usepackage[T1]{fontenc} 
\usepackage{graphicx}
\usepackage[range-phrase={\text{~to~}},output-decimal-marker={.}]{siunitx}
\usepackage[colorlinks,linkcolor=black,urlcolor=black,citecolor=black]{hyperref}
\usepackage{cleveref}
\usepackage[colorinlistoftodos,prependcaption,textsize=tiny]{todonotes}
\usepackage{xcolor}

\DeclareSIUnit\molar{\textsc{M}}
\DeclareSIUnit\photons{\textrm{photons}}
\DeclareSIUnit\pixel{\textrm{pixel}}

\begin{document}

\title{Nanoscale Rigidity in Cross-Linked Micelle Networks Revealed by XPCS Nanorheology}

\author{M.~Reiser}
\email{mario.reiser@xfel.eu}
\affiliation{European XFEL, Holzkoppel 4, 22869 Schenefeld, Germany}
\affiliation{Department Physik, Universität  Siegen, Walter-Flex-Straße 3, 57072 Siegen, Germany}
\author{J.~Hallmann}
\author{J.~Möller}
\author{K.~Kazarian}
\affiliation{European XFEL, Holzkoppel 4, 22869 Schenefeld, Germany}
\author{D.~Orsi}
\affiliation{Department of Mathematical, Physical and Computer Sciences,
  University of Parma, Parco Area Scienze 7/A, 43124 Parma, Italy}
\author{L.~Randolph}
\author{H.~Rahmann}
\affiliation{Department Physik, Universität  Siegen, Walter-Flex-Straße 3, 57072 Siegen, Germany}
\author{F.~Westermeier}
\author{E.~Stellamanns}
\author{M.~Sprung}
\affiliation{Deutsches Elektronen-Synchrotron, Notkestraße 85, 22607 Hamburg, Germany}
\author{F.~Zontone}
\affiliation{European Synchrotron Radiation Facility, 71, avenue des Martyrs, 38043 Grenoble, France}
\author{L.~Cristofolini}
\affiliation{Department of Mathematical, Physical and Computer Sciences,
  University of Parma, Parco Area Scienze 7/A, 43124 Parma, Italy}
\author{C.~Gutt}
\affiliation{Department Physik, Universität  Siegen, Walter-Flex-Straße 3, 57072 Siegen, Germany}
\author{A.~Madsen}
\affiliation{European XFEL, Holzkoppel 4, 22869 Schenefeld, Germany}

\begin{abstract}
  Solutions of wormlike micelles can form cross-linked networks on microscopic
  length scales. The unique mechanical properties of these complex fluids are
  driven by the interplay between the network structure and dynamics which are
  investigated by plate-plate rheometry and X-ray photon correlation
  spectroscopy~(XPCS) nanorheology. Intensity auto-correlation functions of
  tracer nanoparticles~(NPs) dispersed in micelle solutions were recorded which
  captured both the slow structural network relaxation and the short-time
  dynamics of NPs trapped in the network. The results are indicative of a
  resonance-like dynamic behavior of the network on the nanoscale that develops
  as a consequence of the intrinsic short-range rigidity of individual micelle
  chains.
\end{abstract}

\maketitle

Complex networks formed by long cylindrical or wormlike molecules are ubiquitous
in nature and technology. Prominent examples are for instance hydrogels and the
cytoskeleton. These complex networks are mainly functionalized by non-linear
viscoelastic properties where, {\em e.g.} in the
cytoskeleton~\cite{pritchard_mechanics_2014,fletcher_cell_2010}, networks built
up from microscopic filaments determine intracellular transport, mobility and
stabilization~\cite{otten_local_2012,mattila_filopodia_2008,gardel_chapter_2008}
and shear induced strain-stiffening preserves the network shape under external
stresses~\cite{jansen_role_2018}.

Different phenomena have been discussed in terms of their contributions to
non-linear elasticity in complex networks, {\em e.g.}
connectivity~\cite{sharma_strain-controlled_2016,sharma_strain-driven_2016},
bending and stretching forces~\cite{rens_nonlinear_2016}, cross-link
density~\cite{gardel_elastic_2004}, cross-link dynamics and
polydispersity~\cite{meng_fluidization_2018}. Many of these phenomena are
depending on the interplay between the nanoscale network structure and dynamics
where systematic investigations are challenging for both theory and experiments.

A prominent example of such nanoscale cross-linked network systems are wormlike
micelles~\cite{cates_statics_1990,dreiss_wormlike_2007}. Above a critical
concentration, surfactant molecules self-assemble to form micelles. The micellar
shape is determined by properties of the surfactant molecule and in particular
by the ratio of the volumes of its hydrophobic and hydrophilic parts. Wormlike
micelles are promising candidates in the pursue of new ``smart'' materials that
change their function and structure in response to external stimuli (electrical,
optical, thermal,
etc.)~\cite{matsumoto_photo_2008,chu_smart_2013-1,feng_smart_2015}.

\begin{figure}[htbp]
  \centering
  \includegraphics[trim={0 4mm 0 0}]{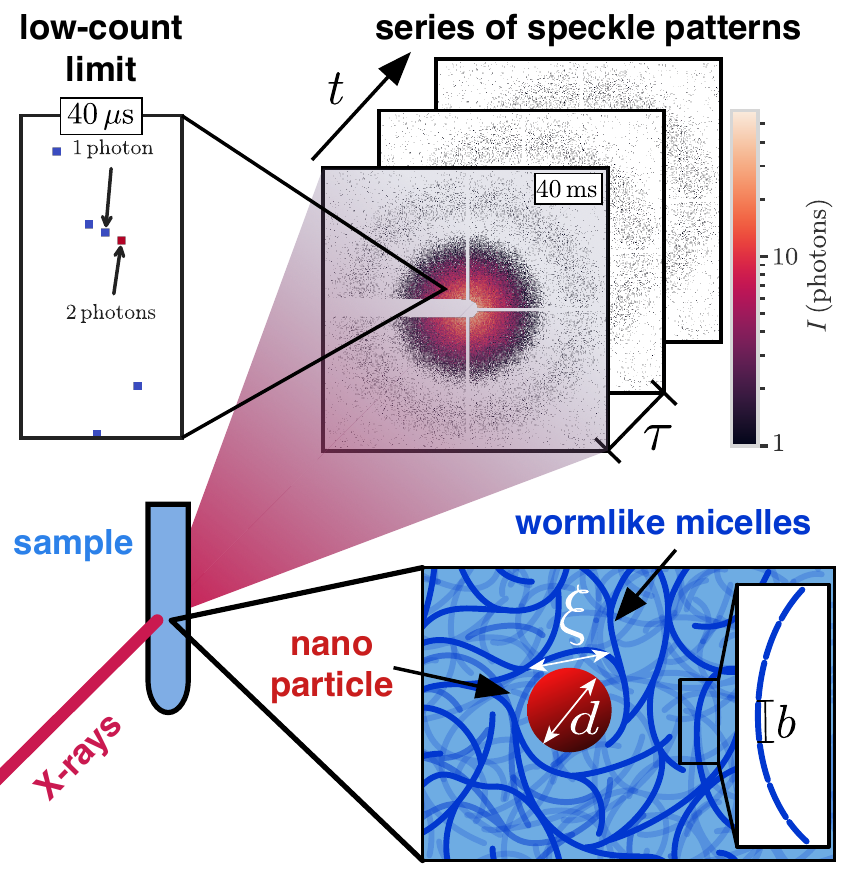}
  \caption[Experiment]{Experimental schematics: Nanoparticles with a diameter of
    ${d=\SI{100}{\nm}}$ are dispersed in the OMCA-CTAB solutions. The
    characteristic mesh size of the network is denoted $\xi$ and $b$ is the
    Kuhn-length. The sample is illuminated by partially coherent X-rays and
    series of speckle patterns are acquired by a 2D pixel detector in the
    far-field. Speckle series were recorded with different exposure times down
    to the limit of a few photons per frame.}
  \label{fig:experiment}
\end{figure}

\textcite{ketner_simple_2007} discovered that aqueous solutions of cetyl
trimethylammonium bromide~(CTAB)~\cite{cates_statics_1990,tao_effect_2013} and
ortho-methoxycinnamic
acid~(OMCA)~\cite{cohen_photochemistry_1975,atkinson_situ_2004} exhibit
remarkable viscoelastic properties due to the self-assembly of long
(${\simeq\SI{300}{\nm}}$) wormlike micelles that form a cross-linked or branched
network. A characteristic length scale of the network is $\xi$, the hydrodynamic
correlation length or network mesh size~(Fig.~\ref{fig:experiment}). Although
wormlike micelles can be described as flexible chains they possess short-range
rigidity described by the \emph{Kuhn}-length,
$b$~\cite{chen_incorporating_2006}, which is a measure of the chain link
dimension (Fig.~\ref{fig:experiment}).

\begin{figure*}[htbp]
  \centering
  \includegraphics[trim={0 .4cm 0 0}]{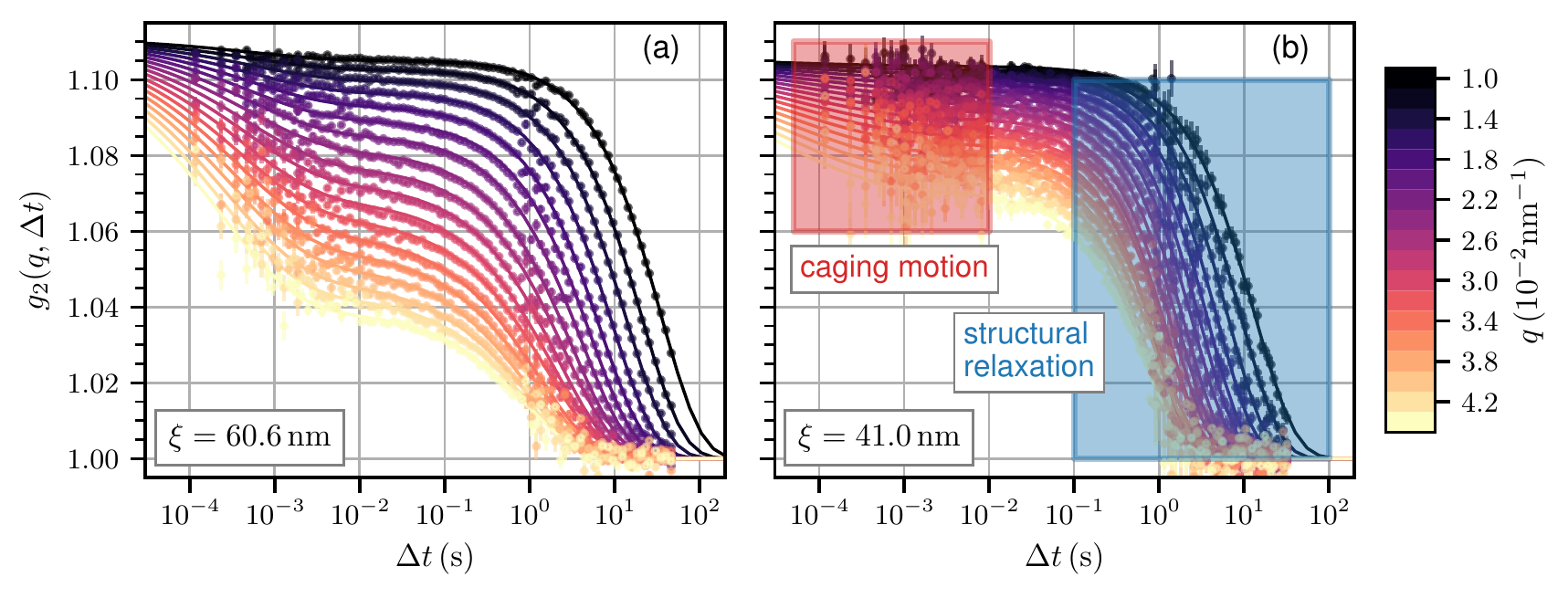}
  \caption[correlation functions]{Intensity auto-correlation functions,
    $g_2(q,\Delta t)$, measured with \SI{100}{\nm} silica NPs dispersed in
    OMCA-CTAB solutions of \SI{50}{\milli\molar}~(a) and
    \SI{80}{\milli\molar}~(b) surfactant and acid concentration. The two
    relaxation modes are indicated in blue and red. Solid lines are fits with a
    double exponential model in Eq.~(\ref{eq:double-exponential}).}
  \label{fig:correlation-functions}
\end{figure*}

Experimentally it is challenging to investigate dynamics on the length scale of
individual chains. Nanorheological techniques hold promise to gain insight into
local phenomena in complex fluids by measuring the dynamics of
nanoparticles~(NPs) dispersed as tracers in the
solutions~\cite{gardel_microrheology_2003,mason_optical_1995,guo_entanglement-controlled_2012}.
One possibility of obtaining dynamical information on the NP dynamics is photon
correlation spectroscopy where the tracer dynamics are deduced from speckle
fluctuations---interference fringes that occur in the far field when the sample
is illuminated by (partially) coherent light. In order to measure NP dynamics in
dense micelle solutions, X-ray photon correlation
spectroscopy~(XPCS)~\cite{grubel_correlation_2004,gutt_measuring_2009,leheny_xpcs_2012,madsen_structural_2016}
was performed~(Fig.~\ref{fig:experiment}). XPCS allows investigation of length
scale dependent dynamics on microscopic length scale due to the short X-ray
wavelength. Novel X-ray detectors are capable of acquiring scattering images
with several kilohertz repetition rate so the full dynamical spectrum from
microseconds to hundreds of seconds is
covered~(Fig.~\ref{fig:correlation-functions}).

OMCA-CTAB samples were prepared according to the recipe of
\textcite{ketner_simple_2007} with surfactant concentrations ranging from
\SIrange{20}{200}{\milli\molar}. The OMCA concentration was equal to the CTAB
concentration for all solutions. Aqueous solutions of OMCA and CTAB were
prepared in separate beakers. When combining the solutions, the viscosity
instantaneously increases due to the formation of the branched wormlike micelle
network. Spherical silica NPs with a diameter of ${d=\SI{100}{\nm}}$ were used
as tracers. The NP volume fraction (\SI{0.04}{\percent}) was tuned to optimize
the scattering signal while reducing possible effects on the micelle formation
to a minimum. The solutions were filled into quartz capillaries with an outer
diameter of \SI{2}{\mm} for the XPCS measurements.

XPCS experiments were conducted at P10 at PETRA~III~(DESY) and ID10~(ESRF)
employing partially coherent X-rays with a photon energy of \SI{8.1}{\keV}. A
Si~(111) monochromator reduced the bandwidth to ${\Delta E/E\approx\num{e-4}}$
to increase the longitudinal coherence. The experiments at ID10 where conducted
with a beam size of \SI{10}{\micro\metre} and a sample-detector distance of
\SI{5.1}{\m}. At P10 the beam size was increased to \SI{75}{\micro\metre} in
favor of a lower radiation dose. In this case the sample-detector distance must
be increased to have an adequate speckle contrast and \SI{20}{\m} was used.
Two-time correlation functions were employed to identify the onset of radiation
damage yielding a dose threshold of \SI{2}{\kilo\gray} which corresponds to less
than one second illumination by the unattenuated beam. Consequently, for every
measurement data were acquired until this dose was reached. Afterwards, the beam
position on the sample was changed for the next acquisition. Attenuating the
beam allowed for including longer time scales. Thereby, the correlation
functions could be measured step-wise and stitched together covering time scales
from microseconds to tens of seconds~(Fig.\ref{fig:correlation-functions}).

The correlation functions exhibit a two-step relaxation process indicative of
the structural network dynamics on long time scales and the localized caging
motion of trapped NPs on short time scales. Hence, the correlation functions can
be modeled by the sum of two exponential modes:
\begin{equation}
  \label{eq:double-exponential}
  g_2(q,\Delta t)=1+\beta_0\sum_{i=1}^2\,\beta_i(q)
  \exp{\left\{-2\left(\Gamma_i(q)\Delta t\right)^{\alpha_i(q)}\right\}}\,,
\end{equation}
where $\beta_0$ is the speckle contrast, $\beta_i$ are the relative strengths
(${\beta_1=1-\beta_2}$), $\Gamma_i$ are the relaxation rates, and $\alpha_i$ are
the Kohlrausch-Williams-Watts~(KWW) exponents. The indices~$1$ and~$2$ are
referring to the short- and long-time relaxation,
respectively~(Fig.~\ref{fig:correlation-functions}).

The localization length, $r_{loc}$, of the NPs is calculated by fitting the
relative strength (non-ergodicity level) as a function of $q$ with
${\beta_2=\exp{(-r_{loc}^2q^2/3)}}$~\cite{leheny_xpcs_2012}. $r_{loc}$ is
related to the hydrodynamic correlation length or network mesh size, $\xi$,
by~\cite{cai_mobility_2011}
\begin{equation}
  \label{eq:correlation-length}
  \xi^3=r_{loc}^2d=k_BT/G_0\,.
\end{equation}
This allows comparing XPCS and classical rheology where $\xi$ is inferred from
the plateau modulus $G_0$. $k_B$ is the Boltzmann constant and $T$ the
temperature.

\begin{figure}[htbp]
  \centering
  \includegraphics[trim={0 .4cm 0 0}]{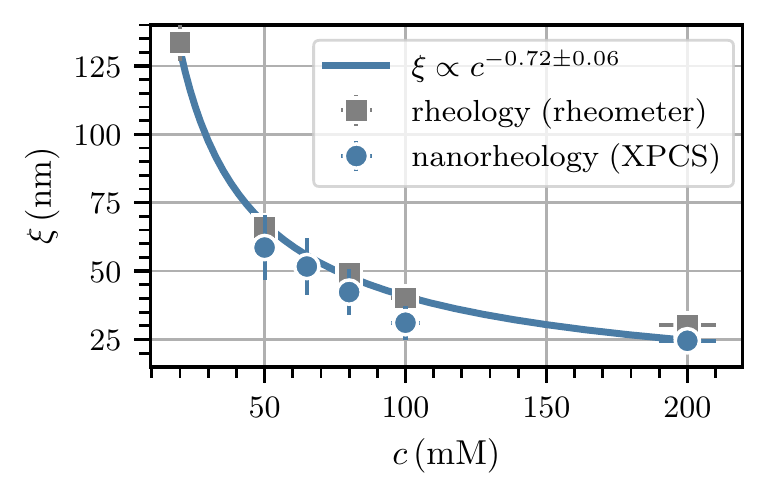}
  \caption{Hydrodynamic correlation length, $\xi$, determined by XPCS based
    nanorheology and classical rheology with a plate-plate rheometer
    (see~Eq.~(\ref{eq:correlation-length}). The solid line indicates a power law
    fit.}
  \label{fig:hydrodynamic-correlation-length}
\end{figure}

Rheology measurements were conducted with a HAAKE MARS~III rheometer at DESY and
$G_0$ was obtained from the complex moduli by fitting a Maxwell model. The
resulting $\xi$ values calculated using Eq.~(\ref{eq:correlation-length}) are
displayed in Fig.~\ref{fig:hydrodynamic-correlation-length} as a function of
concentration,~$c$. Several correlation functions were measured with each
concentration ($\sim$\num{5}~per concentration). The errorbars in
Fig.~\ref{fig:hydrodynamic-correlation-length} reflect the standard deviation of
$\xi$ for different samples with the same concentration. The error of an
individual XPCS measurement is: $\delta\xi\lesssim\SI{2}{\nm}$. Apparently, the
characteristic network length scales determined by both methods (XPCS and
rheology) are in agreement. A power law fit results in a scaling exponent of
\num{-0.72(6)} as expected for a network of semi-flexible wormlike
micelles~\cite{cates_dynamics_1988}.

\begin{figure}[htbp]
  \centering
  \includegraphics[trim={0 .4cm 0 0}]{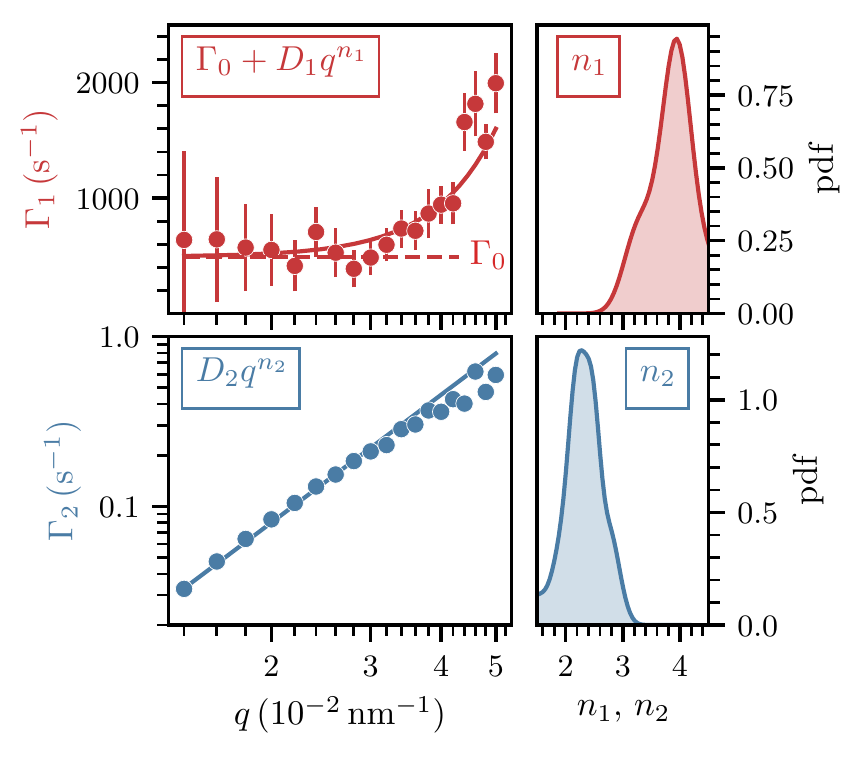}
  \caption[dispersion relation]{(left panels) Dispersion relations of the
    localized caging motion and the structural network relaxation are plotted in
    red and blue, respectively. The data are deduced from the correlation
    function in Fig.\ref{fig:correlation-functions}(a). The functional forms of
    the dispersion relations are indicated on the figures. (right panels)
    Distributions of the exponents, ${n_i}$, over all data sets of correlation
    functions (in total 25).}
  \label{fig:dispersion-relation}
\end{figure}

Information about the dynamics is encoded in the dispersion relations,
$\Gamma_i(q)$, obtained from fits of the correlation functions.
Fig.~\ref{fig:dispersion-relation} displays the relaxation rates obtained from
the data in Fig.~\ref{fig:correlation-functions}(a). The slow structural
relaxation~(blue) is well modeled by a power law of the form
${\Gamma_2(q)=D_2q^{n_2}}$ whereas the short-time dispersion relation,
$\Gamma_1(q)$~(red), can be described by
\begin{equation}
  \label{eq:dispersion-relation}
  \Gamma_1(q)=D_1q^{n_1} + \Gamma_0\,.
\end{equation}
Here, ${D_i}$ are generalized diffusion coefficients and $n_i$ the corresponding
scaling exponents of both relaxation processes. Interestingly, ${\Gamma_1(q)}$
exhibits a clear plateau, $\Gamma_0>0$, for small momentum transfers which is
characteristic of confined dynamics~\cite{bee_quasielastic_1988}.

Correlation functions of different $q$-bins are fitted simultaneously with
Eq.~(\ref{eq:double-exponential})~to~(\ref{eq:dispersion-relation}) to determine
the dynamical parameters. The reader is referred to the SI for a detailed
description of the data treatment. Fig.~\ref{fig:dispersion-relation} displays
histograms of the $q$-scaling exponents, ${n_i}$, over the acquired datasets.
The mean values of the corresponding distributions reflect the different nature
of the two relaxation processes. While the structural relaxation (blue) exhibits
close to diffusive behavior (${n_2=\num{2}}$), the short-time caging motion is
found to be subdiffusive with ${n_1\approx\num{3.9}}$. This subdiffusivity
originates from restricted motion of NPs inside transient network cages and is
hence another evidence for confinement. The KWW~exponents, $\alpha_1$ (not
shown), support the same picture with ${\alpha_1\approx\num{0.5}}$ for the
short-time dynamics meaning that the correlation functions resemble stretched
exponential decays.

\begin{figure}[htbp]
  \centering
  \includegraphics[trim={0 .4cm 0 0}]{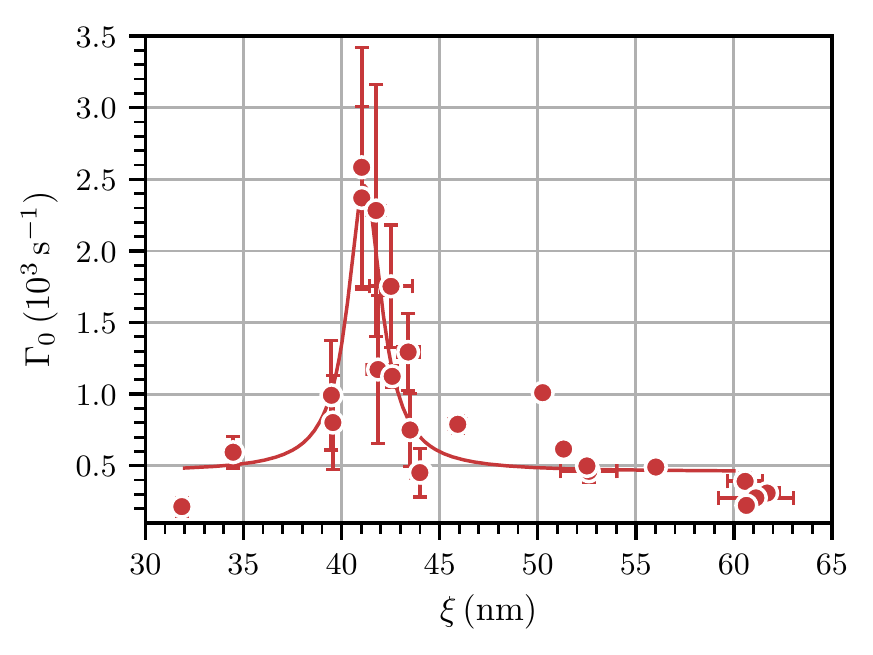}
  \caption[strain stiffening]{The slowest relaxation time under confinement,
    $\Gamma_0$ (plateau value), plotted versus the mesh size, $\xi$ (points).
    The resonance peak of maximum network stiffness is modeled by a Lorentzian
    function plus a constant background (solid line). The peak position is
    ${\xi_0=\SI{41.2(2)}{\nm}}$.}
  \label{fig:nanoscale-rigidity}
\end{figure}

When a NP is trapped inside a network cage, its dynamics are strongly influenced
by interactions with the micelles defining the cage boundaries. Therefore,
microscopic network properties on length scales on the order of the cage size
are related to the short-time NP dynamics which is characterized by the
parameter $\Gamma_0$ introduced in Eq.~(\ref{eq:dispersion-relation}).
$\Gamma_0$ is the slowest relaxation rate of the tracers before they escape
confinement~(structural relaxation) and is hence a measure for the network
stiffness on nanometer length scale. $\Gamma_0$ exhibits a resonance-like
maximum as a function of $\xi$ shown in Fig.~\ref{fig:nanoscale-rigidity}. The
narrow peak occurs at a mesh size of ${\xi_0=\SI{41.2(2)}{\nm}}$ with a full
width at half maximum of \SI{2.1(5)}{\nm} (fitted by a Lorentzian function plus
a constant background of \SI{0.46(8)E3}{\per\second}). This behavior indicates
that the OMCA-CTAB network develops enhanced stiffness resulting in a strong
repulsive force on the NPs leading to an increased relaxation rate. The peak of
$\Gamma_0$ occurs at a hydrodynamic correlation length, $\xi$, close to the Kuhn
length, $b$, found for wormlike CTAB micelles~\cite{chen_incorporating_2006}.
$b$~is the length scale that describes the intrinsic micelle rigidity.
Conclusively, the microscopic network stiffness is determined by the short-range
rigidity of individual micelle strands. Furthermore, the observed maximum of
$\Gamma_0$ shows that the stiffness is non-linearly increasing when the network
mesh size is matching the Kuhn-length of the micelles, similar to a resonance
effect.

The dynamical behavior of the OMCA-CTAB micelle network resembles previously
observed phenomena in branched fiber networks displaying non-linear mechanical
response to
deformation~\cite{sharma_strain-controlled_2016,rens_nonlinear_2016,meng_fluidization_2018,jansen_role_2018}.
Under a certain critical strain, those cross-linked fiber networks undergo a
soft-to-rigid phase transition accompanied by increased susceptibility of
dynamical properties like differential non-affinity~\cite{rens_nonlinear_2016}
or the changing rate of the bending angle~\cite{sharma_strain-driven_2016}.
However, those processes are mostly studied on macroscopic scale using standard
rheometry techniques. Obviously, for understanding fundamental processes in
cross-linked networks and designing new functional nanocomposites, the coupling
of the structural and dynamical network properties has to be studied locally. We
have shown here that nanorheology XPCS is ideally suited to obtain this
information. The XPCS data reveal a connection between the micelle rigidity and
the hydrodynamic network properties in the OMCA-CTAB system on the nanoscale.
The low dose XPCS experiments presented here pave the way for investigations of
general complex networks by studying the interplay between nanoscale structure
and dynamics.

We acknowledge the support from the beam line staff at ID10 and P10 during the
preparation of the experiments and the beamtimes. CG acknowledges support from
BMBF via projects 05K19PS1 and R\"ontgen-\text{\AA}ngstr\"om Cluster Grant
05K20PSA.

\bibliography{./references.bib}
\end{document}


\title{SI: Nanoscale Rigidity in Cross-Linked Micelle Networks Revealed by XPCS Nanorheology}

\author{M.~Reiser, et al.}
\email{mario.reiser@xfel.eu}
\affiliation{European XFEL, Holzkoppel 4, 22869 Schenefeld, Germany}

\maketitle 
\section{Rheology Measurements}
\label{sec:si-rheology}

\begin{figure}[htbp]
  \centering
  \includegraphics[trim={0 4mm 0 0}]{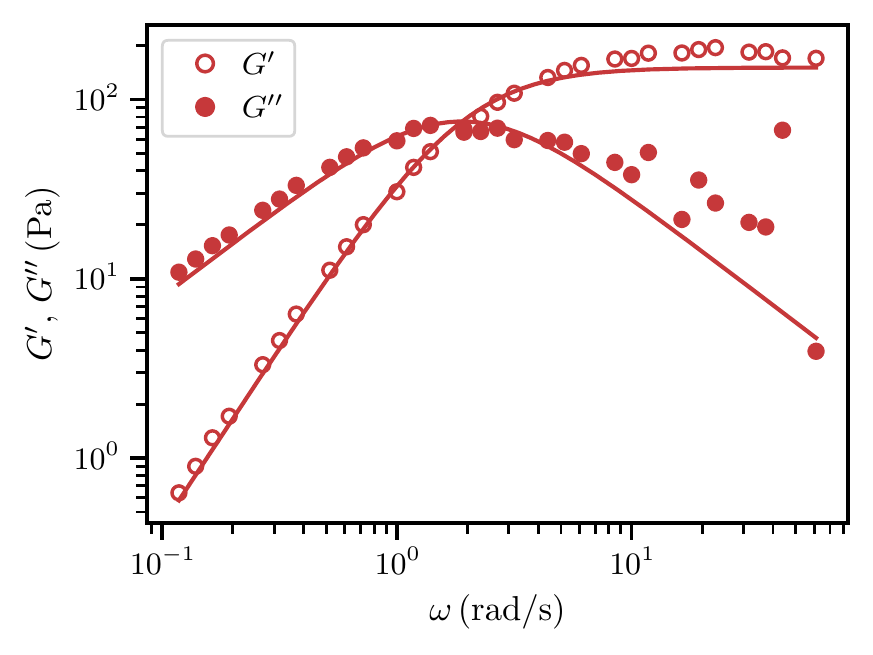}
  \caption{Complex moduli of a \SI{200}{\milli\molar} OMCA-CTAB solution. The
    storage and loss modulus, ${G',\, G''}$, are fitted with the standard
    Maxwell model (solid lines).}
  \label{fig:si-complex-moduli}
\end{figure}

The viscoelastic nature of the OMCA-CTAB micelle solution is evident in the
complex moduli (Fig.~\ref{fig:si-complex-moduli}). The solid line shows a fit
with the standard Maxwell model where the storage and loss modulus as a function
of the shear frequency, $\omega$, are given by
\begin{align}
  \label{eq:storage-maxwell-model}
  G'(\omega) &= G_0\dfrac{\tau_R^2\omega^2}{1+\omega^2\tau_R^2} \
               \quad \text{storage modulus}\,, \\
  \label{eq:loss-maxwell-model}
  G''(\omega) &= G_0\dfrac{\tau_R\omega}{1+\omega^2\tau_R^2} \
                \quad \text{loss modulus}\,,
\end{align}
$\tau_R$ is the terminal stress relaxation time and $G_0$ is the plateau
modulus.

\section{XPCS Measurement Protocol}
\label{sec:si-xpcs-measurement-protocol}

\begin{figure}[htbp]
  \centering
  \includegraphics[trim={0 4mm 0 0}]{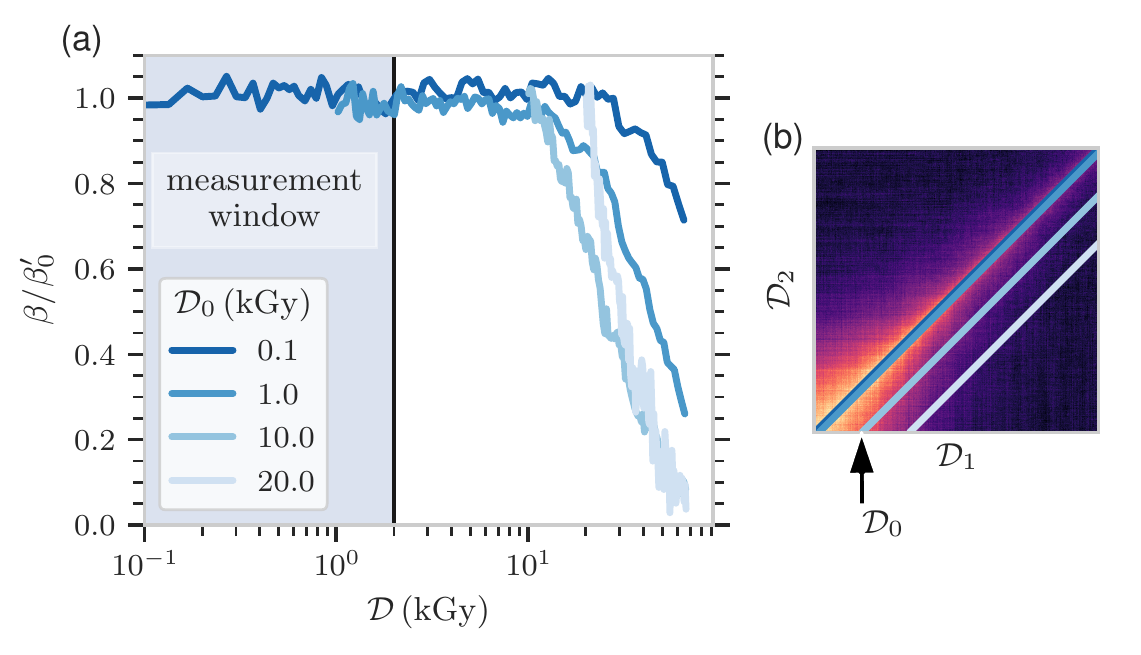}
  \caption{(a) The normalized contrast is extracted from diagonal cuts of the
    time resolved correlation function (b) as a function of the absorbed dose
    during a measurement, ${\mathcal{D}_1,\,\mathcal{D}_2}$. A measurement
    window up to \SI{2}{\kilo\gray} is chosen to safely exclude beam damage
    effects in the XPCS measurements.}
  \label{fig:si-dose-estimation}
\end{figure}

The OMCA-CTAB system is very susceptible to radiation damage and a certain
illuminated volume of the sample can only stand a limited radiation dose before
the properties are altered by the X-rays. To study possible effects of the X-ray
beam on the sample dynamics, time resolved correlation functions are calculated
as a function of the absorbed dose, $\mathcal{D}$, during a measurement:
\begin{equation}
  \label{eq:two-time-correlation}
  c_2(q,\mathcal{D}_1,\mathcal{D}_2)=\frac{\langle I(q,\mathcal{D}_1)I(q,\mathcal{D}_2)\rangle_p}{\langle
    I(q,\mathcal{D}_1)\rangle_p\langle I(q,\mathcal{D}_2)\rangle_p}\,.
\end{equation}
Figure~\ref{fig:si-dose-estimation}(b) displays
$c_2((q,\mathcal{D}_1,\mathcal{D}_2))$ measured with \SI{100}{\nm} silica
nanoparticles in a \SI{50}{\milli\molar} OMCA-CTAB solution. Cuts parallel to
the diagonal (marked in different shades of blue) are plotted as a function of
the absorbed dose in Figure~\ref{fig:si-dose-estimation}(a) and normalized to
the initial contrast. After roughly \SI{10}{\kilo\gray}, the contrast drops
steeply indicating that the X-rays are damaging the micelle network resulting in
an acceleration of the dynamics. To be on the safe side, a maximum dose of
\SI{2}{\kilo\gray} was chosen for the XPCS measurements.

\section{Determination Of Dynamical Parameters}

Fig.~1 in the letter shows that for the shortest exposure time of
\SI{40}{\micro\second} the scattered intensity is reduced to less than
\num{1e-2}~photons per pixel. This requires many repetitions to increase the
signal-to-noise ratio of a set of correlation functions. For instance the
correlation functions shown in Fig.~2 in the main text are calculated from ca.
two million speckle patterns.

Additionally, a global fitting scheme was used for the parameter estimation. The
dynamical parameters $\xi,\,D_i,\,\alpha_i,\,n_i$ and $\Gamma_0$ are calculated
by globally fitting a dataset of correlation functions (all available $q$-bins)
with the double exponential model for ${g_2(q,t)}$ introduced in the letter. The
residuals are minimized by a Levenberg-Marquardt algorithm. The parameters
describing the structural relaxation were considered as free fitting parameters.
Let $N_q$ be the number of $q$-bins, then the parameters of the structural
relaxation are ${\alpha_2^{1}\dots \alpha_2^{N_q}}$ and ${\Gamma_2^{1}\dots
  \Gamma_2^{N_q}}$. The exponent $n_2$ is determined by fitting
${\Gamma_2(q)=D_2q^{n_2}}$. To fit the fast relaxation mode the $q$-dependence
of the short-time dispersion relation is modeled by
${\Gamma_1(q)=\Gamma_0+D_1q^{n_1}}$ resulting in a reduced number of fit
parameters. Consequently, $\Gamma_0$, $D_1$ and $n_1$ are inferred directly from
the correlation functions.

\begin{figure}[htbp]
  \centering
  \includegraphics[trim={0 4mm 0 0}]{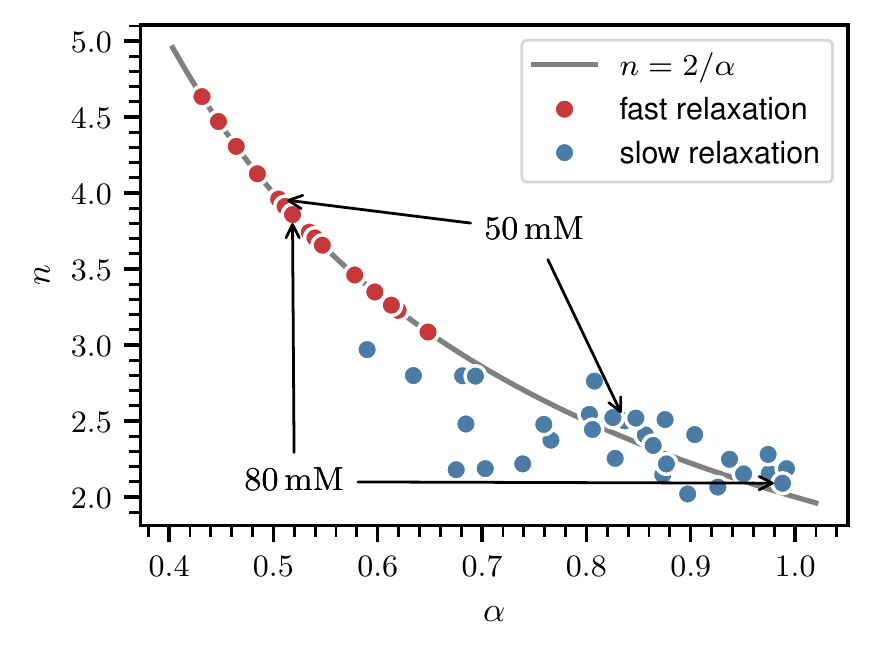}
  \caption{Correlation plot of the $q$-scaling exponent, $n$, and the
    KWW-exponent, $\alpha$. The color distinguishes between the short- and
    long-time relaxation. The solid gray line indicates ${n=2/\alpha}$. The data
    points corresponding to the correlation functions shown in the manuscript
    are annotated.}
  \label{fig:si-correlation-n-kww}
\end{figure}

The parameters $n_2$ and the average $\alpha_2$ are inversely proportional to
each other as shown in Fig.~\ref{fig:si-correlation-n-kww} in blue. Their
relation can be described by
\begin{equation}
  n_i=2/\alpha_i \,,
  \label{eq:n-alpha-relation}
\end{equation}
as indicated by the gray line. In general, the two relaxation processes can only
be resolved with sufficient time resolution and when they are well separated. In
addition to the data shown in the main text where these conditions are fulfilled
Fig.~\ref{fig:si-correlation-n-kww} contains additional datasets where the fast
relaxation process could not be resolved to emphasize the generality of
Eq.~(\ref{eq:n-alpha-relation}). Eq.~(\ref{eq:n-alpha-relation}) is explicitly
used in the fits of the short-time behavior.

\begin{figure}[htbp]
  \centering
  \includegraphics[trim={0 4mm 0 0}]{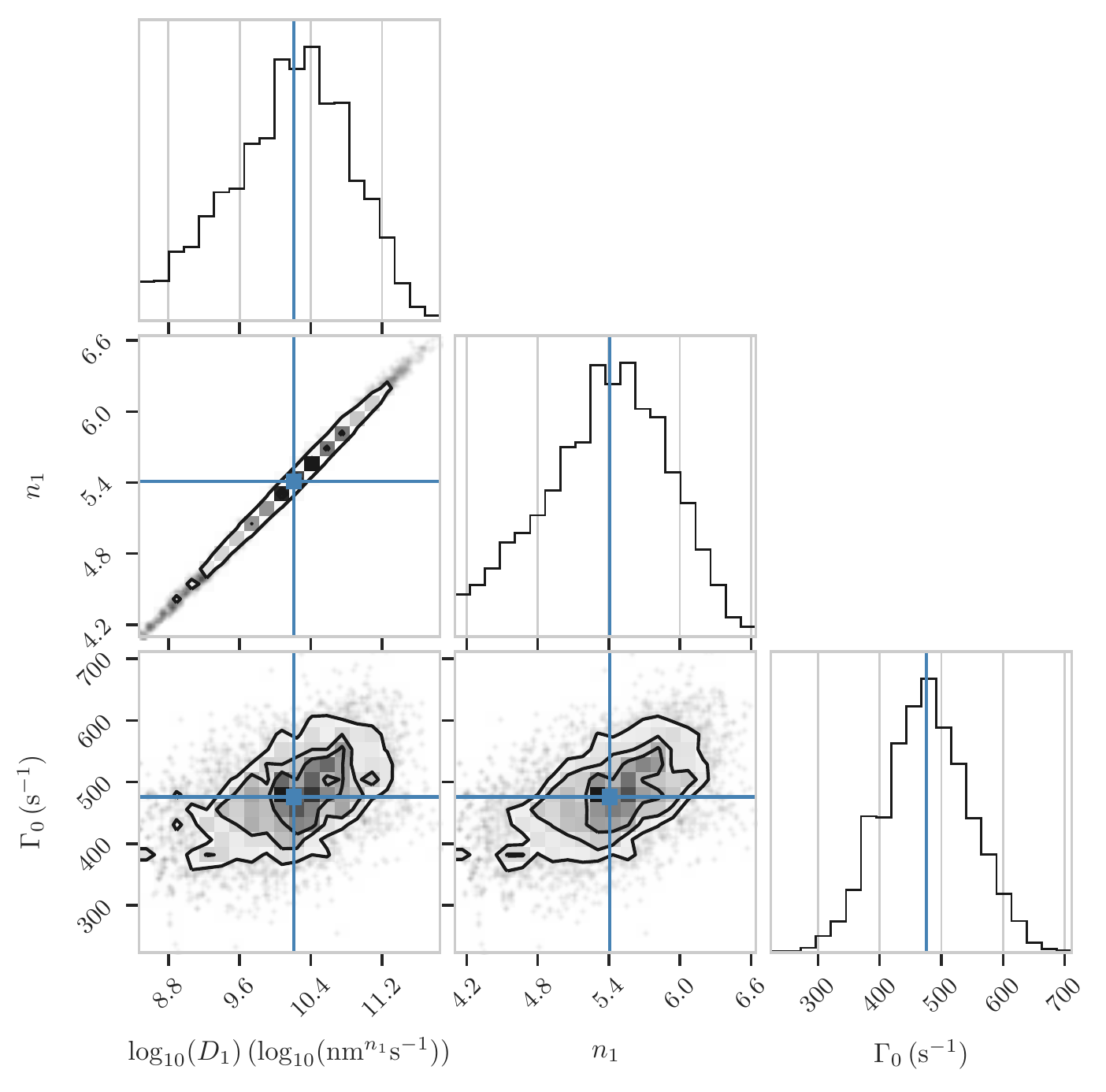}
  \caption{Posterior probability distributions (histograms) and mutual
    correlation plots (others) between fit parameters. The solid lines in the
    correlation plots indicate \SIlist{25;50;75}{\percent}. The blue lines
    indicate the maximum likelihood solution of each parameter. The fits are
    performed with the correlation functions shown in Fig.~2 in the letter.}
  \label{fig:si-corner-plot}
\end{figure}

The reliability of the estimation of $\Gamma_0$ is evident in
Fig.~\ref{fig:si-corner-plot}. The figures show the mutual dependency of the fit
parameters. Clearly, the constant low-$q$ plateau $\Gamma_0$ can be estimated
independently from ${D_1,\,n_1}$ which demonstrates the robustness of the fit.

